\documentclass[11pt,a4paper]{article}

\usepackage{jcappub,bm,subfigure,graphicx}

\title{Permanent mean spin source of the chiral magnetic effect in neutron stars}

\author[a,b]{Maxim Dvornikov}
\author[a]{Victor B. Semikoz}

\affiliation[a]{Pushkov Institute of Terrestrial Magnetism, Ionosphere
and Radiowave Propagation (IZMIRAN),
108840 Moscow, Troitsk, Russia}
\affiliation[b]{Physics Faculty, National Research Tomsk State University,
36 Lenin Avenue, 634050 Tomsk, Russia}
\emailAdd{maxdvo@izmiran.ru}
\emailAdd{semikoz@yandex.ru}

\abstract{
We suggest the generalization of the Anomalous Magneto-Hydro-Dynamics (AMHD) in the chiral plasma of a neutron star (NS) accounting for the mean spin in the ultrarelativistic degenerate electron gas within the magnetized NS core as a continuing source of the chiral magnetic effect. Using the mean field dynamo model generalized in AMHD, one can obtain the growth of a seed magnetic field up to $10^{18}\,\text{G}$ for an old non-superfluid NS at its neutrino cooling era $t < 10^6\,\text{yr}$, while neglecting any matter turbulence within its core and assuming the rigid NS rotation. The application of the suggested approach to the evolution of magnetic fields observed in magnetars, $B\sim 10^{15}\,\text{G}$, should be self-consistent with all approximations used in the suggested laminar dynamo, at least, up to the jumps of growing fields.
}

\keywords{Magnetohydrodynamics, neutron stars, magnetic fields}

\arxivnumber{1904.05768}

\begin{document}

\maketitle

%
%

\section{Introduction}
The axial Ward anomaly known as Adler-Bell-Jackiw  (ABJ) anomaly for the axial-vector current in QED was used to formulate the chiral magnetic effect (CME)
\cite{Vilenkin:1980fu,SonSur09} in different media such as QCD plasma~\cite{Fukushima:2008xe,Zak13} or the hot plasma of the early Universe~\cite{BFR}.

There are multiple attempts to apply the chiral magnetic phenomena in astrophysics. First, we mention recent refs.~\cite{SigLei16,MasKotYam18}, where the Anomalous Mag\-ne\-to-Hyd\-ro-Dy\-na\-mics (AMHD) turbulence is used to generate strong magnetic fields in a protoneutron star (PNS) and explain observed magnetic fields in some compact stars called magnetars~\cite{KasBel17}. The AMHD is applied in ref.~\cite{Kam16} to account for high linear velocities of pulsars. Let us also mark the recent development  on laminar and turbulent dynamos in the AMHD, called the chiral magnetohydrodynamics in ref.~\cite{Rogachevskii:2017uyc}, applied both for the hot plasma in early universe and in PNS in ref.~\cite{Schober:2017cdw}.

A seed magnetic field in PNS can be amplified if the CME is accounted for in the system of chiral fermions. This amplification model is based on the appearance of the chiral imbalance $n_5(t_0)=(n_\mathrm{R} - n_\mathrm{L})\neq 0$ at the initial moment $t_0$, where $n_\mathrm{R,L}(t_0)$ are the initial densities of right and left electrons, which  arises owing to the left electron capture $p + e_\mathrm{L}\to n + \nu_{e\mathrm{L}}$ in a corresponding PNS. Nevertheless the application of the AMHD for the generation of a magnetic field in a neutron star (NS), driven by the CME, fails. The failure of this approach is because of the fast spin-flip due to Coulomb collisions of electrons in NS plasma resulting in $n_5\to 0$ during $\sim 10^{-11}\,\text{s}$~\cite{Dvo16}.

In the present work, we look for alternative mechanisms to support the CME during a time much longer than $10^{-11}\,\text{s}$ accounting both for the chiral vortical effect (CVE)~\cite{Vil79}, which is proportional to the vorticity $\bm{\omega}=\nabla\times \mathbf{v}$, and the chiral separation effect (CSE)~\cite{MetZhi05}, given by the mean spin in magnetized plasma. In particular, we study the saturation
regime, $\partial_t n_5=0$, resulting in $n_5=\text{const}$. It arises due to the CSE when a  magnetic field feeds the chiral imbalance $\mu_5(B) = (\mu_\mathrm{R} - \mu_\mathrm{L})/2 \neq 0$. It happens in the presence of the non-uniform mean spin $\mathbf{S}(\mathbf{x},t)\sim \mu_e(r)\mathbf{B}$, where $\mu_e(r)$ is the electron chemical potential in the NS core.

Then we study the generation of strong magnetic fields driven by the CME in a non-superfluid NS at late stages of its evolution without any turbulence within the NS core.  Note that the turbulence (convection) is rather specific for a nascent NS because of the huge neutrino emission rate $\sim 10^{52}~\text{erg}\cdot \text{s}^{-1}$ leading to the generation of magnetic fields in PNS at the first seconds after a supernova (SN) explosion~\cite{Thompson:1993hn}. 

Thus, instead of the standard MHD dynamo mechanism based on the parameter $\alpha_\mathrm{MHD}$ given by  the kinetic helicity~\cite{Zeldovich}, $\alpha_\mathrm{MHD}\simeq \tfrac{1}{3} \tau_\mathrm{corr}\langle {\bf u}\cdot (\nabla\times\mathbf{u})\rangle$,
where ${\bf u}$ is the random (fluctuation) velocity and $\tau_\mathrm{corr}$ is the correlation time, we apply the CME originated by the presence of the non-uniform chiral imbalance $\mu_5(\mathbf{x},t)\sim \nabla\cdot\mathbf{S}(\mathbf{x},t)$. In this situation, the helicity parameter $\alpha_\mathrm{AMHD}\sim \mu_5(\mathbf{x},t)$ leads to the instability of a seed magnetic field in NS. Note that such a pseudo-scalar field, resulting from the mean spin, has no relation to an effective axion field discussed in ref.~\cite{Boyarsky:2015faa}.

The work is organized as follows. In section~\ref{Sec:AMHD}, we present the full set of AMHD equations completed by the evolution equation for the chiral anomaly density $n_5(\mathbf{x},t)\sim \mu_5(\mathbf{x},t)$. In section~\ref{Sec:master}, we derive the complete system of the evolution equations for the chiral imbalance $\mu_5(\mathbf{x},t)$ and the magnetic field $\mathbf{B}(\mathbf{x},t)$ assuming the rigid NS rotation, $\Omega=\text{const}$. Then, in section~\ref{Sec:CVEvsCME}, we compare the action of  the CME versus the CVE  and then the CSE versus the axial current ${\bf j}_\mathrm{A}$ proportional to the vorticity $\bm{\omega}$.

In section~\ref{Sec:saturation}, we consider the saturation regime $\partial_t\mu_5(\mathbf{x},t)=0$ that allows to reduce the system of the evolution equations in AMHD to the single nonlinear Faraday equation for  $\mathbf{B}(\mathbf{x},t)$. In section~\ref{Sec:alpha_squared}, we present the approximate solution of this Faraday equation using $\alpha^2$-dynamo approach when adopting the magnetic field in the form of the Chern-Simons wave. Then, in section~\ref{Sec:Parker}, we derive  the closed system of the evolution equations for the azimuthal components of the axisymmetric 3D magnetic field: $B_{\varphi}=B(r,\theta,t)$ and the potential $A_{\varphi}=A(r,\theta,t)$. For this purpose we use the mean field dynamo model~\cite{SteKra69} modified here in AMHD. In section~\ref{Sec:diffur}, using the low mode approximation in a thin layer just under the NS crust, $r\lesssim R_\mathrm{NS}$, we derive finally the closed system of the ordinary differential equations for the four amplitudes: $a_{1,2}(t)$ for the azimuthal potential $A$ and $b_{1,2}(t)$ for the toroidal field $B_{\varphi}$. The numerical solution of these equations is illustrated in section~\ref{Sec:Parker1}. Finally, in section~\ref{Sec:Discussion}, we discuss both useful issues and shortcomings of our approach for the generation of strong magnetic fields in such NSs as magnetars.

\section{AMHD in a neutron star with magnetic field \label{Sec:AMHD}}

In this section, we write down the full system of AMHD equations and derive the equation for the magnetic helicity evolution.

The full set of AMHD equations for a plasma in NS, accounting for both the CME and the CVE, is given by~\cite{Sigl,Tashiro:2012mf}
\begin{align}
  \rho
  \left[
    \frac{\partial \mathbf{v}}{\partial t} + (\mathbf{v} \nabla) \mathbf{v} -
    \nu \nabla^2 \mathbf{v}
  \right] = & 
  - \nabla p + {\bf J\times \mathbf{B}},
  \label{Navier-Stokes}
  \displaybreak[2]
  \\
  \frac{\partial \rho}{\partial t} + \nabla (\rho \mathbf{v}) = & 0,
  \label{continuity}
  \\
  \frac{\partial \mathbf{B}}{\partial t} = & - (\nabla \times \mathbf{E}),
  \label{Bianky}
  \displaybreak[2]
  \\
  (\nabla \times \mathbf{B}) = & \sigma 
  \left[
    \mathbf{E} +  \mathbf{v} \times \mathbf{B} + \frac{e^2}{2\pi^2\sigma} \mu_5 \mathbf{B} +
    \frac{e\mu_e}{2\pi^2\sigma}\mu_5\bm{\omega} 
  \right],
  \label{Maxwell}
  \displaybreak[2]
  \\
  \frac{\partial n_5(\mathbf{x},t)}{\partial t} = & - \nabla\cdot \mathbf{S}(\mathbf{x},t) +
  \frac{2\alpha_\mathrm{em}}{\pi}({\bf E}\cdot\mathbf{B}) -\Gamma_f n_5(\mathbf{x},t),
  \label{anomaly_averaged}
\end{align}
where $\mathbf{v}$ is the plasma velocity, $\rho$ is the matter density, $\mathbf{E}$ and $\mathbf{B}$ are the electric and magnetic fields, $p$ is the plasma pressure, $\sigma$ is the electric conductivity, $\nu$ is the viscosity coefficient, $\Gamma_f$ is the spin flip rate, and $\alpha_\mathrm{em}=e^2/4\pi \approx (137)^{-1}$ is the fine structure constant. Here, in contrast, e.g., to refs.~\cite{Sigl,Tashiro:2012mf}, we presented eq.~(\ref{anomaly_averaged}) in its local form and   modified eq.~(\ref{anomaly_averaged}) adding a new pseudoscalar term given by the divergence of the mean spin ${\bf S}(\mathbf{x},t)$.

Equation~(\ref{anomaly_averaged}) stems from the statistical averaging of the ABJ anomaly,
\begin{equation}\label{pseudovector}
  \frac{\partial }{\partial x^{\mu}}\bar{\psi}\gamma^{\mu}\gamma^5\psi =
  2\mathrm{i}m_e\bar{\psi}\gamma_5\psi +
  \frac{e^2}{8\pi^2}F_{\mu\nu}\tilde{F}^{\mu\nu},
\end{equation}
completed by losses for $n_5(\mathbf{x},t)$ due to the spin-flip through collisions in NS plasma ($\sim \Gamma_f$).
In eq.~\eqref{anomaly_averaged}, $n_5(\mathbf{x},t)=\langle\psi^+\gamma_5\psi\rangle=n_\mathrm{R}(\mathbf{x},t) - n_\mathrm{L}(\mathbf{x},t)$ is the chiral anomaly density. The densities of right and left electrons are related to the corresponding chemical potential by $n_\mathrm{R,L}(\mathbf{x},t)=\mu_\mathrm{R,L}^3(\mathbf{x},t)/6\pi^2$. Thus, for $\mu_5\ll \mu_e$ one obtains that $n_5(\mathbf{x},t)=\mu_e^2(r)\mu_5(\mathbf{x},t)/\pi^2$ in the NS degenerate electron gas where $\mu_e(r)$ is the non-uniform chemical potential. Using the results of ref.~\cite{Kelly}, we get that the conductivity of NS matter at the temperature $T=10^{9}\,\text{K}$ has the value $\sigma\simeq 10^7\,\text{MeV}$. 

Finally the new term in kinetic eq.~(\ref{anomaly_averaged}) contains the mean spin in magnetized plasma,
\begin{equation}\label{CSE}
  \mathbf{S}(\mathbf{x},t)=\langle \psi_e^+\bm{\Sigma}\psi_e\rangle_0 = -
  \left(
    \frac{e\mu_e (r) }{2\pi^2}
  \right)
  \mathbf{B}(\mathbf{x},t),
\end{equation}
which arises due to the plasma polarization through the motion of massless electrons and positrons at the main Landau level along the magnetic field; c.f. eq.~(2.2) in ref.~\cite{Dvornikov:2018tsi}. Note that the mean spin in eq.~(\ref{CSE}) does not depend on the plasma temperature. It is known as the CSE (see, e.g., refs.~\cite{MetZhi05,Kharzeev:2015znc}).

Then, integrating local  eq.~(\ref{anomaly_averaged}) over space, $V^{-1}\int {\rm d}^3x(\dots)$, one obtains  the master eq.~(3.14) in ref.~\cite{Dvornikov:2018tsi} for the ultrarelativistic plasma $E/m=\gamma\gg 1$ when $\mathbf{S}_\mathrm{eff}\to \mathbf{S}$\footnote{In ref.~\cite{Dvornikov:2018tsi}, the evolution eq. (\ref{new_Law}) stems from the same ABJ anomaly (\ref{pseudovector}) without the collision term $\sim \Gamma_f$. The densities $n_\mathrm{R,L}(t)=V^{-1}\int   n_\mathrm{R,L}(\mathbf{x},t){\rm d}^3x$, as well as the magnetic helicity density $h(t)$, depend on time only. The sum $\mathbf{S}_\mathrm{eff}=\mathbf{S}_5 + \mathbf{S}$ reduces to the mean spin in eq.~(\ref{CSE}), $\mathbf{S}_\mathrm{eff}\to \mathbf{S}$, since $\mathbf{S}_5$, originated by the mean pseudoscalar $\langle\bar{\psi}_e\gamma_5\psi_e\rangle$, vanishes in the massless limit $m_e=0$, cf. ref.~\cite{Fukushima}.},
\begin{equation}\label{new_Law}
  \frac{{\rm d}}{{\rm d}t}
  \left(
    n_\mathrm{R} - n_\mathrm{L} + \frac{\alpha_\mathrm{em}}{\pi}h
  \right) =
  - \frac{\alpha_\mathrm{em}}{\pi V}\oint_S([{\bf E}\times {\bf A} +
  A_0\mathbf{B}]\cdot{\bf n})\rm{d}^2S - 
  \oint_S
  (\mathbf{S}\cdot {\bf n})
  \frac{d^2S}{V}.
\end{equation}
The magnetic helicity density $h=V^{-1}\int \mathrm{d}^3x({\bf A}\cdot\mathbf{B})$ in eq.~(\ref{new_Law})  evolves as
\begin{equation}\label{standard1}
  \frac{{\rm d}h}{{\rm d}t}= - 2\int \frac{\mathrm{d}^3x}{V}({\bf E}\cdot\mathbf{B})- 
  \oint({\bf n}\cdot\left[\mathbf{B}A_0 +{\bf E}\times {\bf A}\right])\frac{\rm{d}^2S}{V},
\end{equation}
which is known to result in the standard MHD~\cite{Priest}. Note that the evolution of $n_\mathrm{R,L}$ and $h$ in eq.~\eqref{new_Law} is gauge dependent due to the presence of the surface terms.

\section{Evolution of the chiral anomaly in NS \label{Sec:master}}

In this section, we derive the modified Faraday equation and the equation for the evolution of the chiral imbalance accounting for both the CME and the CVE.

Substituting the electric field ${\bf E}$ from the Maxwell eq.~(\ref{Maxwell}) into eq.~(\ref{anomaly_averaged}) and using the chiral imbalance $\mu_5=(\mu_\mathrm{R} -\mu_\mathrm{L})/2=\pi^2n_5/\mu_e^2$, we can change
the term $2\alpha_\mathrm{em}({\bf E}\cdot\mathbf{B})/\pi$ there as
\begin{equation}
  \frac{2\alpha_\mathrm{em}}{\pi}({\bf E}\cdot\mathbf{B})=
  \frac{2\alpha_\mathrm{em}}{\pi}
  \left[
    \frac{\mathbf{B}\cdot(\nabla\times\mathbf{B})}{\sigma} -
    \left(
      \frac{\mu_e^2}{2\sigma}
    \right)
    \left(
      \frac{eB}{\mu_e^2}
    \right)^2
    n_5 -
    \left(
      \frac{e\mu_e}{2\sigma}
    \right)
    \left(
      \frac{\bm{\omega}\cdot\mathbf{B}}{\mu_e^2}
    \right)
    n_5
  \right]. 
\end{equation}
Such a modification allows to recast local eq.~(\ref{anomaly_averaged}) as
\begin{align}\label{recust}
  \frac{\partial n_5(\mathbf{x},t)}{\partial t} = &
  - \nabla\cdot \mathbf{S}(\mathbf{x},t) +
  \frac{2\alpha_\mathrm{em}}{\pi\sigma}[\mathbf{B}\cdot(\nabla\times \mathbf{B})]
  \nonumber
  \\
  &
  -n_5(\mathbf{x}, t)
  \left[
    \Gamma_f + \frac{(2\alpha_\mathrm{em})^2}{\sigma\mu_e^2}B^2 +
    \frac{\alpha_\mathrm{em} e\mu_e}{\pi\sigma\mu_e^2}
    (\bm{\omega}\cdot\mathbf{B})
  \right].
\end{align}
Returning to the chiral imbalance $\mu_5(\mathbf{x},t)=\pi^2n_5(\mathbf{x},t)/\mu_e^2(r)$, one obtains
\begin{equation}\label{anomaly}
  \frac{\partial {\mu_5}(\mathbf{x},t)}{\partial t}=\frac{\pi^2}{\mu_e^2}
  \left\{
    -\nabla\cdot\mathbf{S} +
    \frac{2\alpha_\mathrm{em}}{\pi\sigma}
    \left[
      \mathbf{B}\cdot(\nabla\times\mathbf{B})
    \right]
  \right\} -
  \Gamma_f\mu_5\left[1 + \frac{B^2}{B_0^2} + \frac{\mu_e(\bm{\omega}\cdot\mathbf{B})}{eB_0^2}\right],
\end{equation}
where $B_0^2=\Gamma_f\sigma\mu_e^2/(2\alpha_\mathrm{em})^2=3.1\times 10^4\,\text{MeV}^4$. The strength of this magnetic field is $B_0\simeq 10^{16}\,\text{G}$, which corresponds to $\Gamma_f=10^{11}\,\text{s}^{-1}$. The electric conductivity is $\sigma=10^7\,\text{MeV}$~\cite{Kelly} at the temperature $T=10^9\,\text{K}$ and $\mu_e\simeq 100\,\text{MeV}$ in the NS core\footnote{We put $\mu_e$ stationary and uniform, $\mu_e\simeq 100~{\rm MeV}$, except of a dependence of the quantum (spin) contribution $\sim \mathbf{S}$ on $\mu_e=\mu_e(r)$, given by the NS density profile $n_e(r)=\mu_e^3/3\pi^2$, see in eq. (\ref{CSE}).}.

Substituting the electric field ${\bf E}$ from the Maxwell eq.~(\ref{Maxwell}) into eq.~(\ref{Bianky})
one obtains the induction (Faraday) equation in AMHD,
\begin{equation}\label{Faraday}
  \frac{\partial \mathbf{B}}{\partial t}=
  \nabla\times (\mathbf{v}\times \mathbf{B}) + \frac{1}{\sigma}\nabla^2\mathbf{B} +
  \frac{e}{2\pi^2\sigma}\nabla\times \mu_5(\mathbf{x},t)[e\mathbf{B} + \mu_e\bm{\omega}],
\end{equation}
which completes the system of AMHD eqs.~(\ref{anomaly}) and~(\ref{Faraday}) for the two functions, $\mathbf{B}(\mathbf{x}, t)$ and $\mu_5(\mathbf{x},t)$.

Note that we assume the rigid rotation of NS when the dynamo term $\sim \mathbf{v}$ in the Faraday eq.~(\ref{Faraday}) vanishes. Indeed, since one neglects the turbulent velocity ${\bf u}$, the fluid velocity $\mathbf{v}=\mathbf{V} + {\bf u}$ coincides with the NS rotation $\mathbf{V}=\bm{\Omega}\times {\bf r}$ as a whole. For rigid rotation $\Omega=\text{const}$, the differential rotation is absent, $\partial \Omega/\partial r=0$ and $\partial \Omega/\partial \theta=0$. Thus, the first dynamo term in the Faraday eq.~(\ref{Faraday}) vanishes. In addition, we neglect below a small vorticity term $\sim\bm{\omega}=\nabla\times {\bf v}$ both in eq. (\ref{Faraday}) and in eq. (\ref{anomaly}). The validity of this approximation is discussed in section~\ref{Sec:CVEvsCME}.  For such a case we do not need to involve the Navier-Stokes equation, while the instability of the magnetic field  is presented through the last term in eq.~(\ref{Faraday}). Here we assume also a non-uniform  chiral imbalance $\mu_5(\mathbf{x},t)=\pi^2n_5(\mathbf{x},t)/\mu_e^2(r)$ for the stationary  non-uniform electron density in NS, $n_e(r)=\mu_e^3(r)/3\pi^2=n_cY_e(1 - r^2/R_\mathrm{NS}^2)$~\cite{Lattimer}, where $n_c=10^{38}\,\text{cm}^{-3}$ is the central (neutron) density and $Y_e\simeq 0.04$ is the electron abundance.

The formal solution of eq.~(\ref{anomaly}) without the small vorticity term $\sim \bm{\omega}$ takes the form,
\begin{equation}\label{solution}
  \mu_5(\mathbf{x},t)=\mu_5(t_0)e^{-A(t)} +
  \left(
    \frac{\pi^2}{\mu_e^2}
  \right)
  e^{-A(t)}\int_{t_0}^t\mathrm{d}t' e^{+A(t' )}
  \left[
    -\nabla\cdot\mathbf{S} +
    \frac{2\alpha_\mathrm{em}}{\pi\sigma}\mathbf{B}\cdot(\nabla\times \mathbf{B})
  \right].
\end{equation}
In eq.~\eqref{solution}, the index $A(t)=A(\mathbf{x},t)$ in the exponent reads
\begin{equation}\label{Aindex}
  A(\mathbf{x},t)=\Gamma_f\int_{t_0}^t \mathrm{d}t' 
  \left(
    1 + \frac{B^2(\mathbf{x},t' )}{B_0^2}
  \right), 
\end{equation}
which is huge\footnote{$\Gamma_f=10^{11}\,\text{s}^{-1}=6.6\times 10^{-11}\,\text{MeV}$ is the rate of spin-flip~\cite{Dvo16}.} for times $(t - t_0)\gg 10^{-11}\,\text{s}$. It means that the initial imbalance $\mu_5(t_0)$ vanishes rapidly in eq.~(\ref{solution}). However, the main question in this paper whether $\mu_5(t)$ survives in SN or vanishes at all requires to solve, in a self-consistent way, the Faraday eq.~(\ref{Faraday}), which depends on the mean spin $\mathbf{S}(\mathbf{x},t)$ through the second integral term for $\mu_5(\mathbf{x},t)$ in eq.~(\ref{solution}).

\section{Comparison of the contributions of the CVE and the CME\label{Sec:CVEvsCME}}

In this section, we compare the contributions of the CVE and the CME to the system of AMHD equations.

The total vector current $\mathbf{j}=\mathbf{j}_\mathrm{Ohm} + \mathbf{j}_{\chi}$ in the Maxwell eq.~(\ref{Maxwell}),  where $\mathbf{j}_\mathrm{Ohm}=\sigma({\bf E} + \mathbf{v}\times \mathbf{B})$, includes also the sum of the CME and the CVE~\cite{Tashiro:2012mf}, $\mathbf{j}_{\chi}=\mathbf{j}_{\chi B} + \mathbf{j}_{\chi \omega}$. Here $\mathbf{j}_{\chi B}=e^2\mu_5\mathbf{B}/2\pi^2$ is the anomalous current that drives the CME and $\mathbf{j}_{\chi \omega}=e(\mu_\mathrm{R}^2 - \mu_\mathrm{L}^2)\bm{\omega}/8\pi^2=e\mu_e\mu_5\bm{\omega}/2\pi^2$ is given by the fluid vorticity $\bm{\omega}=\nabla\times \mathbf{v}$~\cite{Vil79}. For $\mathbf{j}_{\chi \omega}$ we substituted $\mu_\mathrm{R} - \mu_\mathrm{L}=2\mu_5$ and $\mu_\mathrm{R} + \mu_\mathrm{L}=2\mu_e$, which stems from the sum $n_e=n_\mathrm{R} + n_\mathrm{L}=\mu_e^3/3\pi^2$, where $n_\mathrm{L,R}=\mu_\mathrm{L,R}^3/6\pi^2$.

Substituting $\bm{\omega}=2\bm{\Omega}$
for the NS rotation as a whole with the velocity $\mathbf{v}=\bm{\Omega}\times {\bf r}$, where $\Omega\simeq 10^3\,\text{s}^{-1}=6.6\times 10^{-19}\,\text{MeV}$  is the NS rotation frequency\footnote{We use units $\hbar=c=1$ for which  $\text{s}^{-1}=6.6\times 10^{-22}\,\text{MeV}$.}, one can compare the values of $j_{\chi B}$ and $j_{\chi \omega}$. We find the CME prevails over the CVE, $j_{\chi B}\gg j_{\chi \omega}$ since $eB\gg 2\mu_e\Omega$. Substituting $\mu_e=100\,\text{MeV}$, one gets
\begin{equation}\label{inequality}
  eB\gg 13.2\times 10^{-17}\,\text{MeV}^2=66\times 10^{-4}\,\text{G},
  \quad
  {\rm or}
  \quad
  B\gg 10^{-2}\,\text{G},
\end{equation}
where the known relations ${\rm MeV}^2=5\times 10^{13}\,\text{G}$ and $e=\sqrt{4\pi\alpha_\mathrm{em}}=0.3$ are used. 

Let  us stress that both $\mathbf{j}_{\chi B}\sim \mu_5\bm{\omega}$ and $\mathbf{j}_{\chi B}\sim \mu_5\mathbf{B}$ are the {\it pure vector currents} since $\mu_5$ is the pseudoscalar, $\mu_5\to - \mu_5$, with respect to the space inversion. There is the additional {\it axial current} given by the vorticity $\bm{\omega}$~\cite{Lan11,Kharzeev:2015znc},
\begin{equation}\label{rotation1}
  \mathbf{j}_\mathrm{A}=
  \left[
    \frac{T^2}{6} + \frac{1}{2\pi^2}(\mu_e^2 + \mu_5^2)
  \right]
  \bm{\omega}\approx \frac{\mu_e^2}{2\pi^2}\bm{\omega},
\end{equation}
which cannot compete with the axial mean spin $\mathbf{S}= -(e\mu_e /2\pi^2) \mathbf{B}$ (the CSE) that contributes to the density imbalance evolution in eq.~(\ref{anomaly_averaged}). Indeed, if we substitute a fast NS rotation frequency $\omega=2\Omega=10^{3}\,\text{s}^{-1}$ into eq.~(\ref{rotation1}), one obtains that $| \mathbf{S}|\gg |\mathbf{j}_\mathrm{A}|$, comparing eqs.~(\ref{CSE}) and~(\ref{rotation1}), due to the same inequality in eq.~(\ref{inequality}), $eB\gg 2\mu_e\Omega$, or  $B\gg 10^{-2}\,\text{G}$.

Thus, the CVE is negligible for NS plasma, as well as the axial current in eq.~(\ref{rotation1}), in comparison with
the mean spin (the CSE) contribution to eq.~(\ref{anomaly_averaged}).

\section{Saturation regime \label{Sec:saturation}}

In this section, we explore the saturation regime of the chiral imbalance in the magnetic field evolution. It should be noted that the saturation of the $\alpha$-parameter in the solar dynamo was discussed previously. It leads to the algebraic quenching~\cite{Cha10}.

From eq.~(\ref{anomaly}) one gets that, in the case $\partial_t\mu_5(\mathbf{x},t)=0$, the stationary $\mu_5^{(\mathrm{sat})}(\mathbf{x})$ reads
\begin{align}\label{saturation}
  \mu_5^{(\mathrm{sat})}(\mathbf{x})= &
  \left(
    \frac{\pi^2}{\mu_e^2\Gamma_f}
  \right)
  \frac{[(2\alpha_\mathrm{em}/\pi\sigma)\mathbf{B}\cdot(\nabla\times\mathbf{B}) -
  \nabla\cdot\mathbf{S}]}{1 + B^2/B_0^2 +\mu_e(\bm{\omega}\cdot\mathbf{B})/eB_0^2}
  \nonumber
  \\
  & \approx
  \left(
    \frac{\pi}{2\alpha_\mathrm{em} B_0^2}
  \right)
  \left[
    \mathbf{B}\cdot (\nabla\times \mathbf{B}) -
    \left(
      \frac{\pi\sigma}{2\alpha_\mathrm{em}}
    \right)
    \nabla\cdot\mathbf{S}
  \right],
\end{align}
where we use the inequalities $B_0^2=\Gamma_f\mu_e^2\sigma/(2\alpha_\mathrm{em})^2\gg B^2\gg \mu_e\omega B$ in the denominator of the first line in eq.~(\ref{saturation}) neglecting there the vorticity term $\sim \bm{\omega}$. Of course, we assume here that the magnetic field reaches the saturation so that $\partial_t\mathbf{B}=0$. We expect this to happen before $\mu_5^{(\mathrm{sat})}$ settles.

For the negative derivative $\mathrm{d}n_e/\mathrm{d}r = n_cY_e(-2r/R_\mathrm{NS}^2)<0$, the term $-\nabla\cdot \mathbf{S}$ in eq.~(\ref{saturation}) equals to 
\begin{equation}\label{spinterm}
  -\nabla\cdot \mathbf{S}= + \frac{eB_r}{2\pi^2}\frac{{\rm d}\mu_e(r)}{{\rm d}r}=
  \frac{eB_r}{2\mu_e^2}\frac{{\rm d}n_e(r)}{{\rm d}r}\simeq - 10^{-17}
  \left(
    \frac{r}{R_\mathrm{NS}}
  \right)
  \left(
    \frac{B_r}{{\rm MeV}^2}
  \right)
  {\rm MeV}^4,
\end{equation}
where $B_r$ is the radial component of the magnetic field. In eq.~\eqref{spinterm}, we substitute the mean spin $\mathbf{S}$ from eq.~(\ref{CSE}) and use the electron density profile\footnote{We substitute  $n_cY_e=8\times 10^{5}\,\text{MeV}^3\times 0.04$ in eq.(\ref{spinterm}), where $n_c=n_n=10^{38}~{\rm cm}^{-3}=8\times 10^{5}\,\text{MeV}^3$ is the central (neutron) density and $Y_e=0.04$
 is the electron abundance, as well as put $R_\mathrm{NS}^{-1}=2\times 10^{-17}\,\text{MeV}$ for $R_\mathrm{NS}= 10^6\,\text{cm}$.} $n_e(r)=\mu_e^3(r)/3\pi^2=n_cY_e(1 - r^2/R_\mathrm{NS}^2)$, proposed in ref.~\cite{Lattimer}.

Substituting the spin term $(\nabla\cdot\mathbf{S})$ from eq.~(\ref{spinterm}) into eq.~(\ref{saturation}), one can estimate the value $\mu_5^{(\mathrm{sat})}(r,\theta)$ just below the NS crust, $r\lesssim R_\mathrm{NS}$,
\begin{align}\label{saturation2}
  \frac{\mu_5^{(\mathrm{sat})}(R_\mathrm{NS},\theta)}{{\rm eV}}=&
  \frac{\pi^2\sigma}{{\rm eV}\,(2\alpha_\mathrm{em})^2B_0^2}
  \left[
    \frac{e B_r(R_\mathrm{NS},\theta)}{2\mu_e^2}\frac{{\rm d}n_e}{{\rm d}r} +
    \frac{2\alpha_\mathrm{em}}{\pi\sigma}\mathbf{B}\cdot(\nabla\times \mathbf{B})
  \right]
  \nonumber
  \\ &
  = -  3\times 10^{-4}
  \left(
    \frac{B_r(R_\mathrm{NS},\theta)}{{\rm MeV}^2}
  \right) +
  1.4\times 10^{-15}R_\mathrm{NS}
  \left(
    \frac{\mathbf{B}\cdot(\nabla\times \mathbf{B})}{{\rm MeV}^4}
  \right).
\end{align}
Note that the non-uniform $\mu_5^{(\mathrm{sat})}(r,\theta)$ remains parity odd with respect to the space inversion analogously to $\mu_5(t)$ in a uniform matter: $\mu_5\to - \mu_5$, since components $B_i(r,\theta)$, $i=r,\theta,\varphi$, like the diffusion term in eq.~\eqref{saturation2}, are pseudoscalars for the total axial vector $\mathbf{B}=\sum_i B_i\hat{\bf e}_i$, which is parity even accounting for the unit vectors $\hat{\bf e}_i$, $\hat{\bf e}_i\hat{\bf e}_k=\delta_{ik}$. In what follows, we neglect the small diffusion term in eq.~(\ref{saturation2}) since it does not change issues in the dynamo model~\cite{Parker,SteKra69} applied below in section~\ref{Sec:Parker}.

\subsection{Is the anomaly saturation in eq.~(\ref{saturation2}) sufficient to drive the growth of the magnetic field? \label{Sec:alpha_squared}}

First, we mention that it is very difficult to solve the dynamical system of eqs.~(\ref{anomaly}) and~(\ref{Faraday}). Assuming instead the saturation solution of eq.~(\ref{anomaly}) given by eq.~(\ref{saturation2}), where we omit the diffusion term, and substituting such $\mu_5^{(\mathrm{sat})}(B_r)$ 
into Faraday eq.~(\ref{Faraday}) for the rigid NS rotation, one gets the nonlinear Faraday equation,
\begin{equation}\label{Faraday1}
  \frac{\partial \mathbf{B}}{\partial t}=\frac{1}{\sigma}\nabla^2\mathbf{B} +
  \frac{2\alpha_\mathrm{em}}{\pi\sigma}
  \nabla\times \mu_5^{(\mathrm{sat})}(B_r)\mathbf{B},
\end{equation}
where $B_r$ should vary somehow together with $\mu_5^{(\mathrm{sat})}$.

In order to feel how efficient is the instability driven by the modified $\mu_5$, we could assume changing (probe) radial components $B_r(R_\mathrm{NS})/{\rm MeV}^2=0.02\div 20$, or $B_r(R_\mathrm{NS})=10^{12}\,\text{G}\div 10^{15}\,\text{G}$ independently of the latitude. This could be implemented, e.g., for the magnetic field which depends on one spatial coordinate,  $\mathbf{B}(z,t)=(\pm \sin kz, \cos kz, 0)e^{\zeta t}$~\cite{Zeldovich}, where $\zeta=-\eta k^2 \pm \alpha(t)k$, that is similar to the Chern-Simons wave. The simplified solution of eq.~(\ref{Faraday1}) rewritten in the standard MHD form,
\begin{equation}
  \frac{\partial \mathbf{B}}{\partial t}=\eta \nabla^2\mathbf{B} +
  \alpha (t)\nabla\times \mathbf{B},
\end{equation}
where $\eta=\sigma^{-1}$, $\alpha (t) =2\alpha_\mathrm{em}\mu_5^{(\mathrm{sat})}(B_r)/\pi\sigma$ for changing $B_r$, takes the following form:
\begin{equation}\label{dynamo}
  B(t,k,B_r)=B_0\exp
  \left[
    \int_{t_0}^t(\pm \alpha (B_r)  k - \eta k^2)\mathrm{d}t'
  \right]
  \Longrightarrow B_0\exp
  \left[
    \frac{\alpha^2(B_r)}{4\eta}(t - t_0)
  \right]. 
\end{equation}
Such a solution corresponds to the $\alpha^2$-dynamo amplification of the initial amplitude $B_0$ valid for the extreme wave number $k_\mathrm{ext}=|\alpha|/2\eta$ in the interval $|\alpha|/\eta> k\geq k_\mathrm{ext}$ where the field in eq.~(\ref{dynamo}) is growing.

One can easily find  from eq.~(\ref{saturation2}) the corresponding wave number
\begin{equation}
  k_\mathrm{ext}=
  \left(
    \frac{| B_r|}{{\rm MeV}^2}
  \right)
  \frac{1}{14\,{\rm cm}}.
\end{equation}
The scale $\lambda_\mathrm{ext}=14\,{\rm cm}/(| (B_r|/{\rm MeV}^2)$ diminishes from $7~{\rm m}$ at $B_r=10^{12}\,\text{G}$ down to $\lambda_\mathrm{ext}=0.7\,{\rm cm}$ for growing $B_r=10^{15}\,\text{G}$. The index $\alpha^2(t - t_0)/4\eta$ in the exponent, entering eq.~(\ref{dynamo}), becomes bigger than unity at times $t> 4\times 10^4\,\text{yr}$ for the magnetic field $| B_r|/{\rm MeV}^2=1$ close to the Schwinger value, and for times $t> 100\,\text{yr}$ if the field rises up to $B_r=10^{15}\,\text{G}$ observed in magnetars~\cite{KasBel17}. 

For extreme $B_r=10^{16}\,\text{G}$, the  $\alpha^2$-dynamo mechanism is efficient starting even from the early NS age $t> 1\,\text{yr}$. Of course, for $B_r\to B_0=10^{16}\,\text{G}$ we should include the saturation factor $(1 + B^2/B_0^2)^{-1}$ presented in the first line in eq.~(\ref{saturation}).

\subsection{Mean field dynamo for the 3D axially symmetric field \label{Sec:Parker}}

To proceed in the analysis of eq.~\eqref{Faraday1}, we use the spherical coordinates system, that is natural if one deals with a magnetic field in NS. Then, following ref.~\cite[pg.~373]{Sti04}, we decompose the magnetic field into the toroidal $\mathbf{B}_t = B_\varphi \mathbf{e}_\varphi$ and the poloidal $\mathbf{B}_p = B_r \mathbf{e}_r + B_\theta \mathbf{e}_\theta$ components: $\mathbf{B} = \mathbf{B}_t + \mathbf{B}_p$. Moreover, we introduce the vector potential $\mathbf{A}= A_\varphi \mathbf{e}_\varphi$ for $\mathbf{B}_p = (\nabla \times \mathbf{A})$.

Using in eq.~(\ref{Faraday1}) the helicity parameter $\alpha_\mathrm{sat}=2\alpha_\mathrm{em}\mu_5^{(\mathrm{sat})}(\mathbf{B})/\pi\sigma$ given by the chiral imbalance $\mu_5^{(\mathrm{sat})}(\mathbf{B})$ in eq.~(\ref{saturation2}), one can get for the toroidal component $B\equiv B_{\varphi}$ the following nonlinear equation:
\begin{align}\label{dynamo1}
  \frac{\partial B}{\partial t}= & \eta
  \left(
    \nabla^2 B -\frac{B}{r^2\sin^2\theta}
  \right) +
  \frac{1}{r}\frac{\partial}{\partial r}(r\alpha_\mathrm{sat}B_{\theta}) -
  \frac{1}{r}\frac{\partial}{\partial \theta}(\alpha_\mathrm{sat}B_r)
  \nonumber
  \\
  & = 
  \eta
  \left(
    \frac{1}{r}\frac{\partial^2(rB)}{\partial r^2} + \frac{1}{r^2}\frac{\partial}{\partial \theta}
    \left[
      \frac{1}{\sin \theta}\frac{\partial (\sin \theta B)}{\partial \theta}
    \right]
  \right)
  \nonumber
  \\
  & -
  \frac{1}{r}\frac{\partial}{\partial r}
  \left[
    \alpha_\mathrm{sat}\frac{\partial}{\partial r}(rA)
  \right] -
  \frac{1}{r}\frac{\partial}{\partial \theta}
  \left[
    \frac{\alpha_\mathrm{sat}}{r\sin\theta}\frac{\partial}{\partial \theta}(\sin \theta A)
  \right],
\end{align}
where $\eta=\sigma^{-1}=\beta$ is the magnetic diffusion coefficient. Here we do not take into account the turbulence contribution $\beta_\mathrm{T}$ to $\beta$; cf. ref.~\cite[pg.~370]{Sti04}. 
In the last line of eq.~\eqref {dynamo1}, we substitute the poloidal components $B_{r,\theta}$ for the axisymmetric field given by the azimuthal potential $A\equiv A_{\varphi}$,
\begin{equation}\label{components}
  B_{\theta}=- \frac{1}{r}\frac{\partial}{\partial r}(rA),
  \quad
  B_{r}=\frac{1}{r\sin\theta}\frac{\partial}{\partial \theta}(\sin\theta A),
\end{equation}
where $A = A(r,\theta,t)$.

Finally the dynamo in AMHD is completed by the equation for the azimuthal potential $A$,
\begin{equation}\label{dynamo2}
  \frac{\partial A}{\partial t}= \eta
  \left(
    \frac{1}{r}\frac{\partial^2(rA)}{\partial r^2} + \frac{1}{r^2}\frac{\partial}{\partial \theta}  
    \left[
      \frac{1}{\sin \theta}\frac{\partial (\sin \theta A)}{\partial \theta}
    \right]
  \right) +
  \alpha_\mathrm{sat}B.
\end{equation}
Equations~\eqref{dynamo1} and~\eqref{dynamo2} should be further analyzed and solved numerically.

\subsection{Low mode approximation in the thin layer $r\simeq R_\mathrm{NS}$ \label{Sec:diffur}}

To solve eqs.~\eqref{dynamo1} and~\eqref{dynamo2} we have to sepatate the variables $\theta$ and $t$. For this purpose we follow the approach of ref.~~\cite{Sokoloff} and represent the functions $A$ and $B$ in the Fourier series,
\begin{equation}\label{lowmode}
  A(t,\theta)=a_1(t)\sin\theta  + a_2(t)\sin 3\theta +\dotsc,
  \quad
  B(t,\theta)=b_1(t)\sin 2\theta + b_2(t)\sin 4\theta + \dotsc,
\end{equation}
where the orthogonal functions obey the relations,
\begin{equation}
  \int_0^{\pi}\mathrm{d}\theta  \sin \theta \sin 3\theta =
  \int_0^{\pi}\mathrm{d}\theta \sin 2\theta \sin 4\theta =0,
\end{equation}
and normalized as
\begin{equation}
  \frac{2}{\pi}\int_0^{\pi}\mathrm{d}\theta \sin^2 (l \theta) =1,
  \quad
  l = 1,2,\dots.
\end{equation}
Neglecting the radial dependence in a thin layer close to the NS crust, $r\lesssim R_\mathrm{NS}$, analogously to ref.~\cite{Parker}, we simulate the derivative with respect to $r$ as $R_\mathrm{NS}\mathrm{d}/\mathrm{d}r\to \mathrm{i}\mu$. Thus the parameter $\mu$ simulates a layer width where the magnetic field changes significantly.

Then, we define the dimensionless time $\tau=t/t_\mathrm{diff}$, where the diffusion time $t_\mathrm{diff}=\sigma_0 R_\mathrm{NS}^2=5\times 10^{11}\,\text{yr}$ exceeds the age of the Universe $t_\mathrm{U}=1.4\times 10^{10}\,\text{yr}$ because of the huge electric conductivity in NS,  
$\sigma_0=10^{7}\,\text{MeV}$, at the temperature $T=10^9\,\text{K}$. Note that the conductivity depends on the temperature $\sigma (T)=\sigma_0(T/T_9)^{-5/3}$~\cite{Kelly}, where $T_9=10^9\,\text{K}$. NS cools down as~\cite{Page:2004fy},
\begin{equation}\label{cooling}
  \frac{T}{T_9}=\left(\frac{t}{t_9}\right)^{-1/6},
\end{equation}
where $t_9$ is the time after the SN explosion when the NS temperature drops down to $T_9$. Hence the conductivity changes over time as $\sigma\equiv\sigma (t)=\sigma_0F$ where $F=(t/t_9)^{0.28}$ and $t=t_9 + \tau\sigma_0R_\mathrm{NS}^2$.
Using figure~7 in ref.~\cite{Gnedin:2000me} for NS with the mass $M=1.7M_{\odot}$, one finds that NS cools down to the temperature $T_9=10^9\,\text{K}$ during $t_9\sim 1\,\text{yr}$.

On this way, multiplying consistently eq.~(\ref{dynamo2}) by $R^2_\mathrm{NS}\eta^{-1}=R^2_\mathrm{NS}\sigma (t)=t_\mathrm{diff}F$ and $F^{-1}$, substituting $\alpha_\mathrm{sat}=2\alpha_\mathrm{em}\mu_5^{(\mathrm{sat})}(\mathbf{B})/\pi\sigma (t)$ given by eq.~(\ref{saturation2}) without diffusion term, we can rewrite eq.~(\ref{dynamo2}) as
\begin{align}\label{dynamo22}
  \frac{\partial A}{\partial \tau}= &
  F^{-1}
  \bigg\{
    -\mu^2A + \frac{\partial^2A}{\partial \theta^2} - \frac{A}{\sin^2\theta} +
    \cot \theta \frac{\partial A}{\partial \theta}
    \nonumber
    \\
    & -
    \frac{3\times 10^{7}\alpha_\mathrm{em}}{\pi}
    \left[
      \frac{\partial A}{\partial\theta} + \cot \theta A
    \right]
    \left(
      \frac{B}{{\rm MeV}^2}
    \right)
  \bigg\},
\end{align}
where, in the last term, we substitute the number ${\rm eV}\times R_\mathrm{NS}=10^{11}/2$ since ${\rm eV}=10^5/2~~ {\rm cm}^{-1}$ in units $\hbar=c=1$. Note that this nonlinear term in eq.~(\ref{dynamo22}) is produced by the mean spin term $\nabla\cdot\mathbf{S}$. It is new compared to that we deal with in the standard MHD.

Substituting the decomposition in eq.~(\ref{lowmode}), multiplying by $\sin \theta$ or $\sin 3\theta$, and then integrating over the colatitude $\theta$, one obtains the system in the low mode dynamo approach,
\begin{align}\label{potential}
  \dot{a}_1 = &
  - \frac{1}{F}
  \bigg\{
    (\mu^2 + 2)a_1 + 2a_2
    \notag
    \\
    & +
    \frac{2\kappa[(a_1 - a_2)b_1 + 2a_2b_2 ]}
    {1+K[(b_1^2 + b_2^2) + \mu^2(a_1^2+a_2^2) + 4(a_1^2 + 5a_2^2 + 2a_1 a_2)]/F}
  \bigg\},
  \notag
  \\
  \dot{a}_2= &
  - \frac{1}{F}
  \bigg\{
    (\mu^2 + 12)a_2 
    \notag
    \\
    & +
    \frac{2\kappa [(a_1 + a_2)(b_1 + b_2)]}
    {1+K[(b_1^2 + b_2^2) + \mu^2(a_1^2+a_2^2) + 4(a_1^2 + 5a_2^2 + 2a_1 a_2)]/F}
  \bigg\}, 
\end{align}
where $\kappa=3\times 10^7\alpha_\mathrm{em}/2\pi=3.5\times 10^4$, $K = 1.3\times 10^{-5}$, and the azimuthal field components $b_{1,2}$ are normalized on ${\rm MeV}^2=5\times 10^{13}\,\text{G}$. Note that, in eq.~\eqref{potential}, we account for the quenching factor $(1+B^2/B_0^2)^{-1}$ in eq.~\eqref{saturation}.

Analogously, multiplying consistently the Faraday eq.~(\ref{dynamo1}) by $R_\mathrm{NS}^2\sigma(t)=t_\mathrm{diff}F$ and $F^{-1}$, substituting $\alpha_\mathrm{sat}=2\alpha_\mathrm{em}\mu_5^{(\mathrm{sat})}(\mathbf{B})/\pi\sigma (t)$, given by eq.~(\ref{saturation2}), we rewrite eq.~(\ref{dynamo1}) as
\begin{align}\label{dynamo11}
  \left(\frac{1}{\rm{MeV}^2}\right)\frac{\partial B}{d\tau}= & F^{-1}
  \bigg\{
    \left(\frac{1}{\rm{MeV}^2}\right)\left[ -\mu^2B + \frac{\partial^2B}{\partial \theta^2} + \cot\theta \frac{\partial B}{\partial \theta} - \frac{B}{\sin^2\theta}\right]\nonumber\\ & + 2\kappa \left(\frac{\partial A}{\partial\theta} + \cot\theta A\right)\left[ - \mu^2A + 2\left(\frac{\partial^2A}{\partial \theta^2} + \cot\theta \frac{\partial A}{\partial \theta} - \frac{A}{\sin^2\theta}\right)\right]  
  \bigg\},
\end{align}
where the azimuthal potential $A$ is normalized on $R_\mathrm{NS}\times {\rm MeV}^2$. Thus the components
$a_{1,2}$ in the decomposition in eq.~(\ref {lowmode}) are dimensionless as $A$ there. Let us stress again that the second line $\sim \kappa$ results from the mean spin contribution $\nabla\cdot\mathbf{S}$.

Multiplying eq.(\ref{dynamo11}) by $\sin 2\theta$ or $\sin 4\theta$, integrating over the colatitude $\theta$, we complete the system in eq.~(\ref{potential}) by 
\begin{align}\label{magfield}
  \dot{b}_1= & 
  - \frac{1}{F}
  \bigg\{
    (\mu^2 + 6)b_1 + 4b_2
    \notag
    \\
    & +
    \frac{2\kappa [a_1^2 (\mu^2 + 4) + 24a_1a_2 + a_2^2(\mu^2 + 20)]}
    {1+K[(b_1^2 + b_2^2) + \mu^2(a_1^2+a_2^2) + 4(a_1^2 + 5a_2^2 + 2a_1 a_2)]/F}
  \bigg\},
  \notag
  \\
  \dot{b}_2= & 
  - \frac{1}{F}
  \bigg\{
    (\mu^2 + 20)b_2 
    \notag
    \\
    & +
    \frac{2\kappa[a_1a_2(3\mu^2 + 32) + a_2^2(\mu^2 + 32)]}
    {1+K[(b_1^2 + b_2^2) + \mu^2(a_1^2+a_2^2) + 4(a_1^2 + 5a_2^2 + 2a_1 a_2)]/F}
  \bigg\},
\end{align}
where take into account the quenching factor $(1+B^2/B_0^2)^{-1}$ analogously to eq.~\eqref{potential}. Note that this quenching factor results from eq.~\eqref{saturation}. Thus, we have the four ordinary differential eqs.~(\ref{potential}) and~(\ref{magfield})  for the four amplitudes $a_{1,2}(\tau)$ and $b_{1,2}(\tau)$. 

\subsubsection{Initial conditions\label{sec:inicond}}

To get the initial conditions for eqs.~(\ref{potential}) and~(\ref{magfield}), we equate the initial normalized components 
\begin{align}
  \frac{B_{\theta}(R_\mathrm{NS},\theta,t=0)}{{\rm MeV}^2}= & -
  \left[
    a_1(0)\sin\theta + a_2(0)\sin3\theta
  \right],
  \notag
  \displaybreak[2]
  \\
  \frac{B_r(R_\mathrm{NS},\theta,t=0)}{{\rm MeV}^2}= &
  2\cos\theta
  \left[
    a_1(0) +a_2(0)(4\cos2\theta - 1)
  \right],
\end{align}
that result from eq.~(\ref{components}) in the low mode approximation in eq.~(\ref{lowmode}), $B_{\theta}(t=0)/{\rm MeV}^2=B_r(t=0)/{\rm MeV}^2=0.2$. Then one can find at the same force line of the poloidal field  $B_{p}=\sqrt{B_{\theta}^2 + B_r^2}$, while at different latitudes when substituting corresponding $\theta=0$ for $B_r(t=0)$ where $B_{\theta}(t)=0$  and $\theta=\pi/2$ for $B_{\theta}(t=0)$ where $B_r(t)=0$, the following algebraic system:
\begin{equation}\label{a_initial}
  a_2(0) - a_1(0)=0.2,
  \quad
  a_1 + 3a_2(0)=0.1.
\end{equation}
The opposite signs of the initial amplitudes resulting from eq.~(\ref{a_initial}), $a_1(0)=-0.125$, $a_2(0)=+ 0.075$, provide the growth of the azimuthal magnetic field components $b_{1,2}$ obeying eq.~(\ref{magfield}). Hence the amplitudes $a_{1,2}$, obeying eq.(\ref{potential}), will grow as well. We choose the same initial condition for the azimuthal components at $r\simeq R_\mathrm{NS}$, $b_{1,2}(0)=0.2$, corresponding, at the beginning, to the toroidal field $B(R_\mathrm{NS},\theta, t=0)=(\sin 2\theta + \sin 4\theta)\times 10^{13}\,\text{G}$ in the low mode approximation in eq.~(\ref{lowmode}).

\section{Growth of the magnetic field in AMHD induced by the mean spin \label{Sec:Parker1}}

\begin{figure}
  \centering
  \subfigure[]
  {\label{1a}
  \includegraphics[scale=.35]{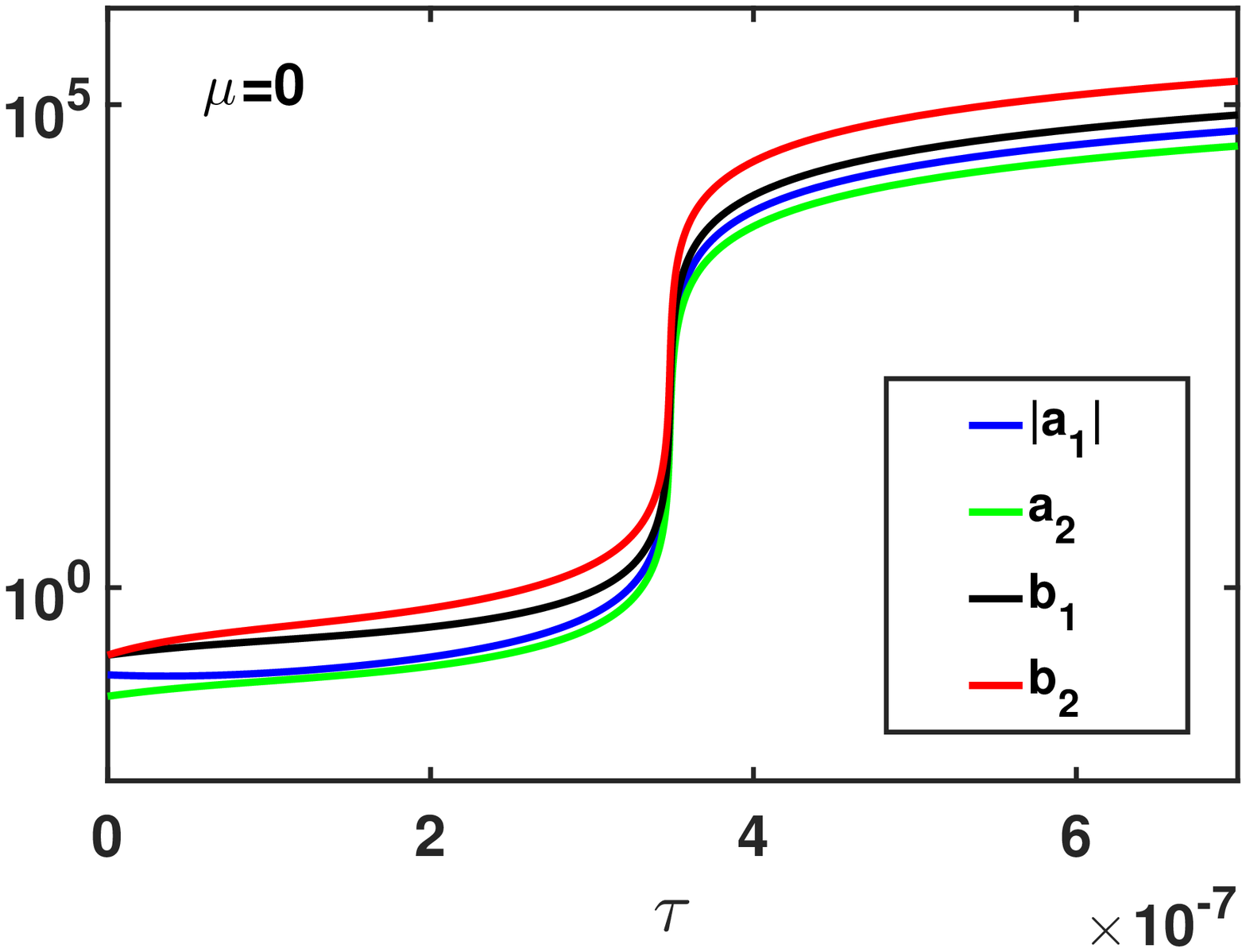}}
  \hskip-.6cm
  \subfigure[]
  {\label{1b}
  \includegraphics[scale=.35]{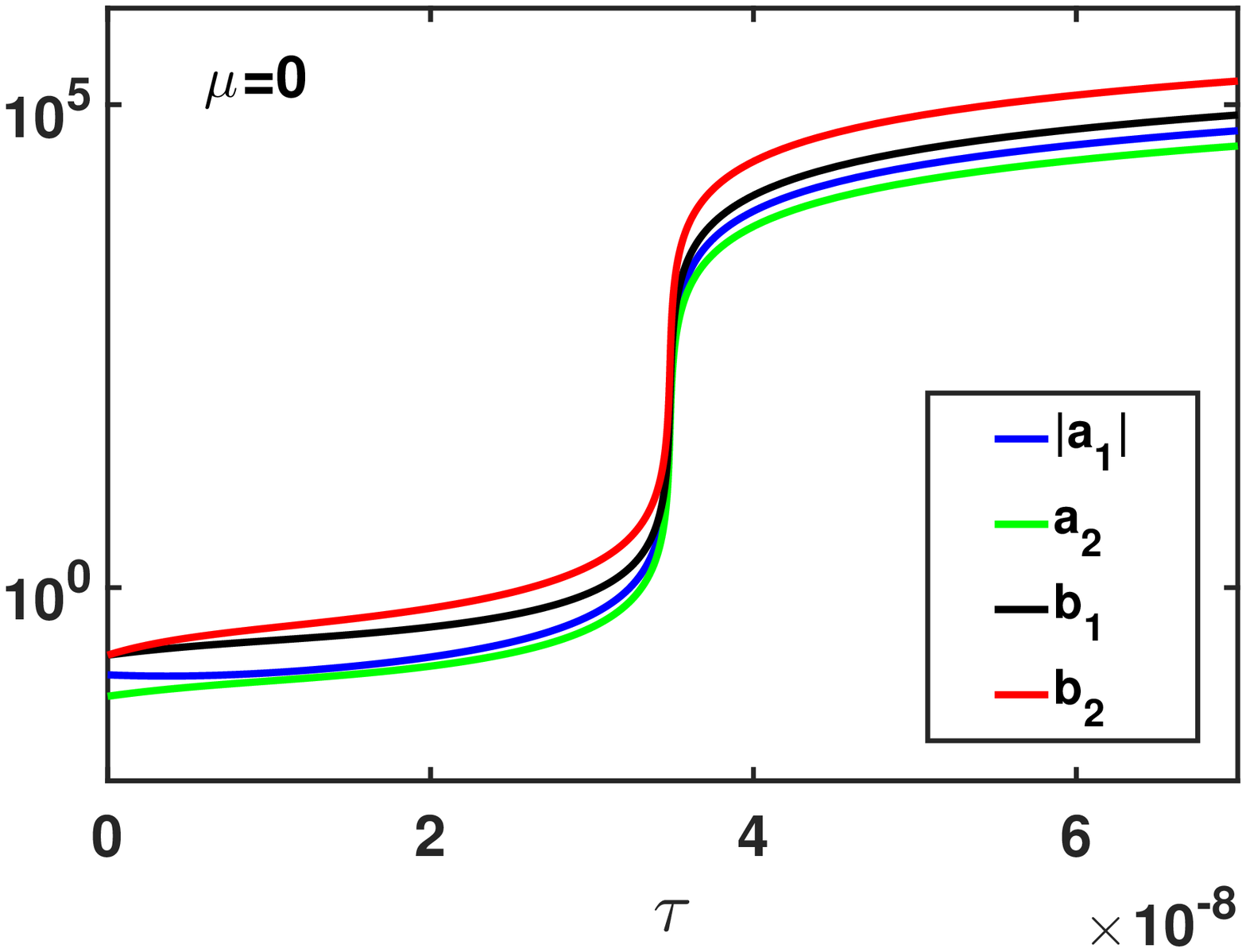}}
  \\
  \subfigure[]
  {\label{1c}
  \includegraphics[scale=.35]{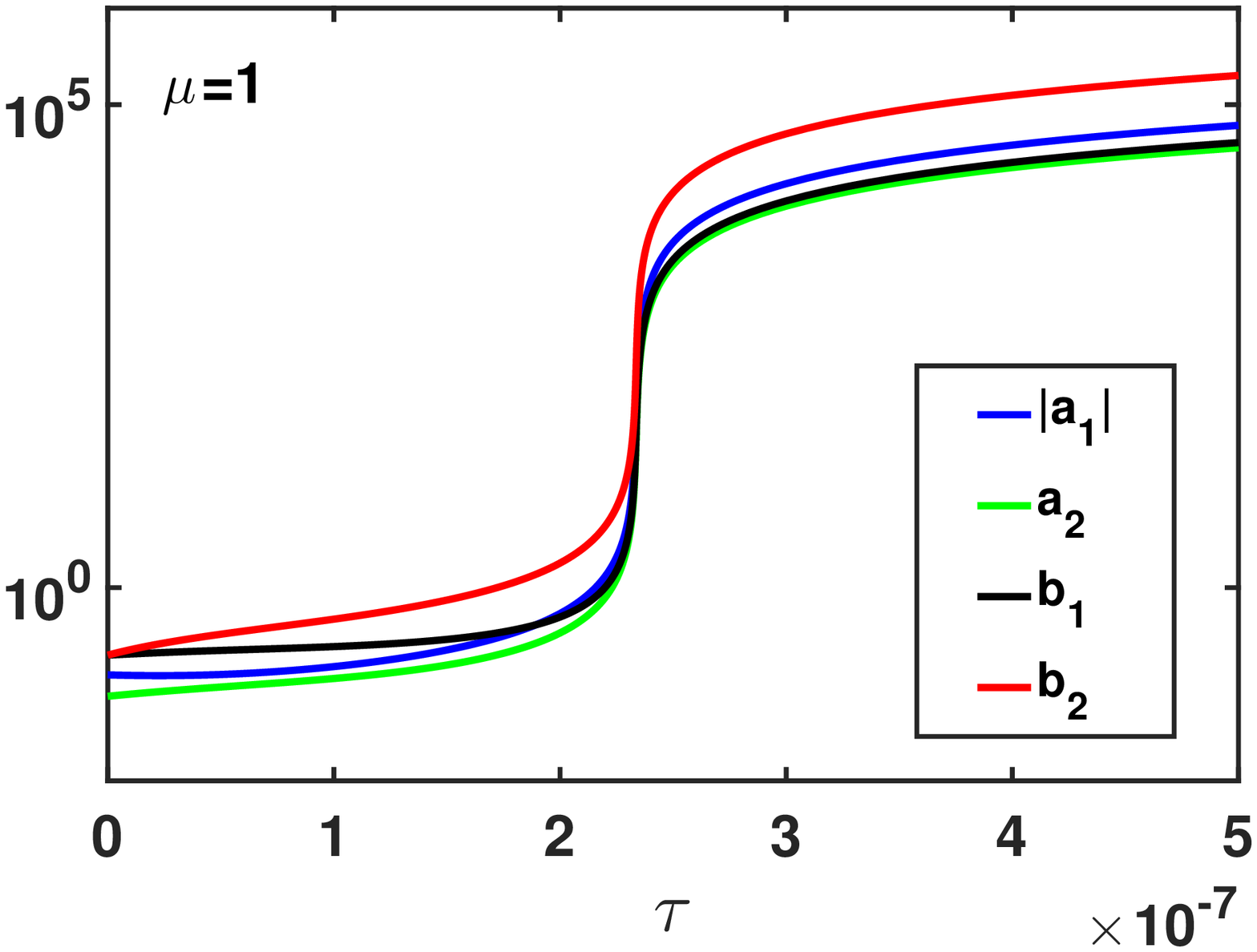}}
  \hskip-.6cm
  \subfigure[]
  {\label{1d}
  \includegraphics[scale=.35]{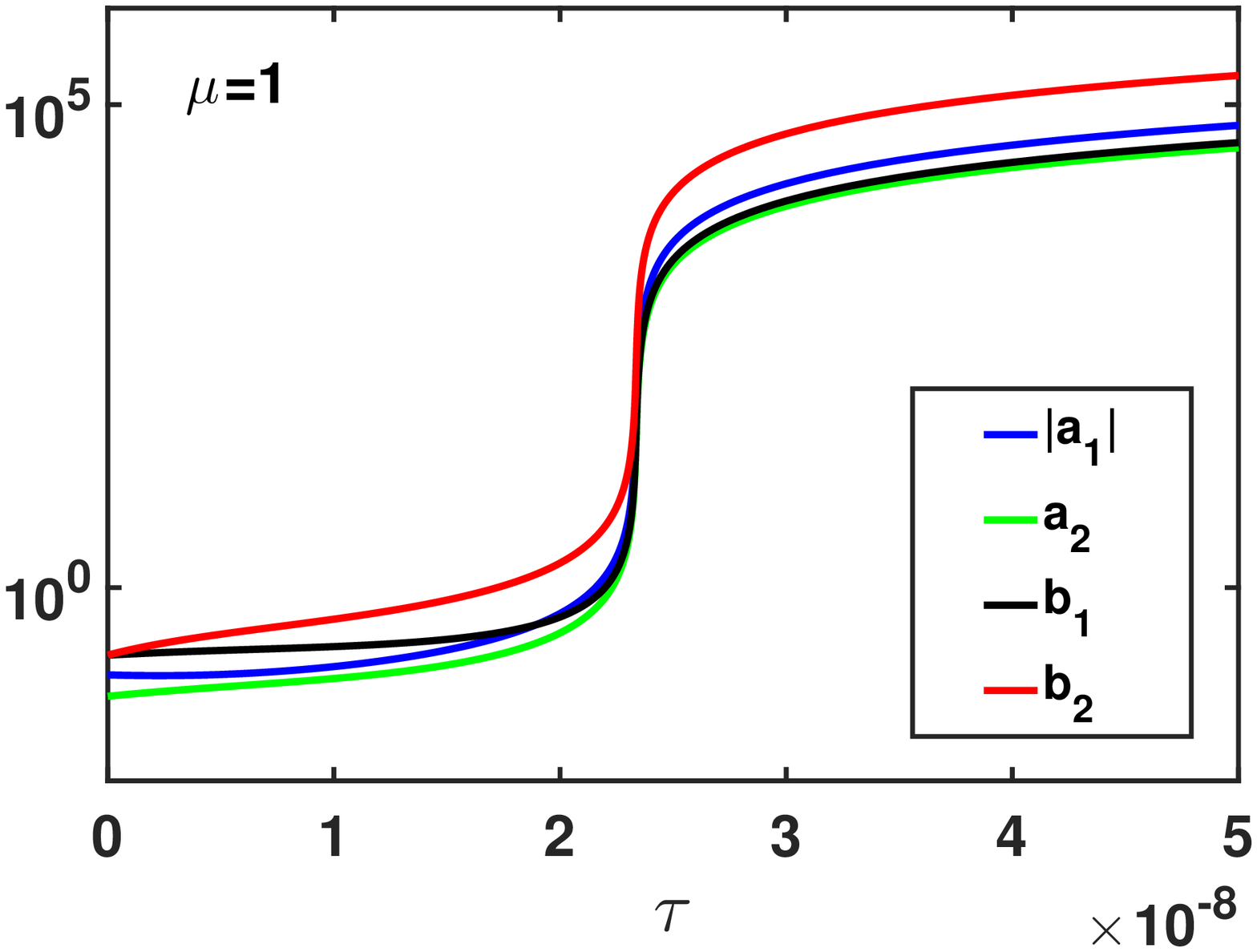}}
  \protect
  \caption{The behavior of the coefficients $a_{1,2}$ and $b_{1,2}$ for different $\mu$
  and $\kappa$ without taking into account the NS cooling, i.e. setting $F = 1$
  in eqs.~(\ref{potential}) and~(\ref{magfield}).
  (a)~$\mu = 0$ and $\kappa = 10^7$;
  (b)~$\mu = 0$ and $\kappa = 10^8$;
  (c)~$\mu = 1$ and $\kappa = 10^7$;
  (d)~$\mu = 1$ and $\kappa = 10^8$.
  \label{fig:nocooling}}
\end{figure}

\begin{figure}
  \centering
  \subfigure[]
  {\label{2a}
  \includegraphics[scale=.35]{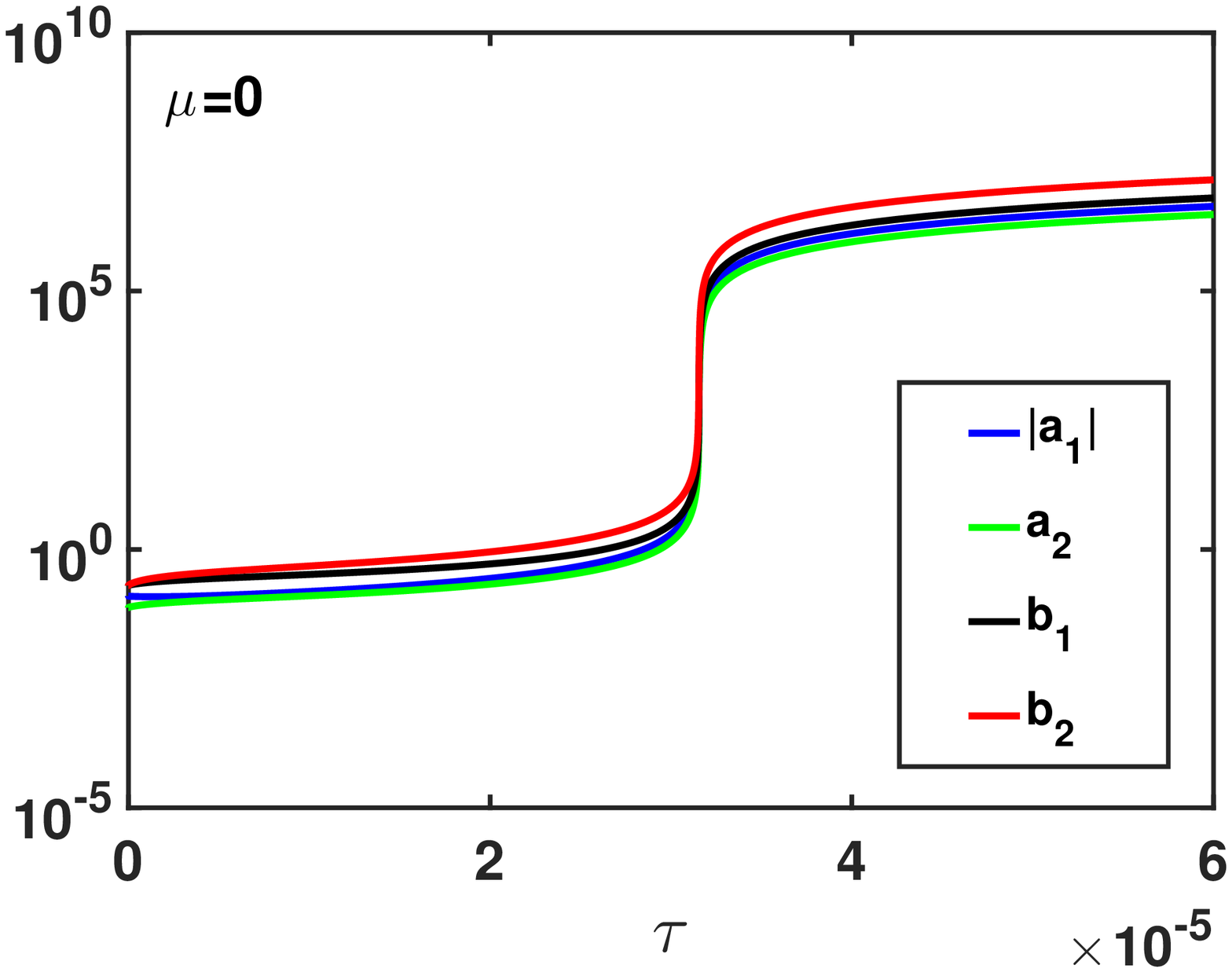}}
  \hskip-.6cm
  \subfigure[]
  {\label{2b}
  \includegraphics[scale=.35]{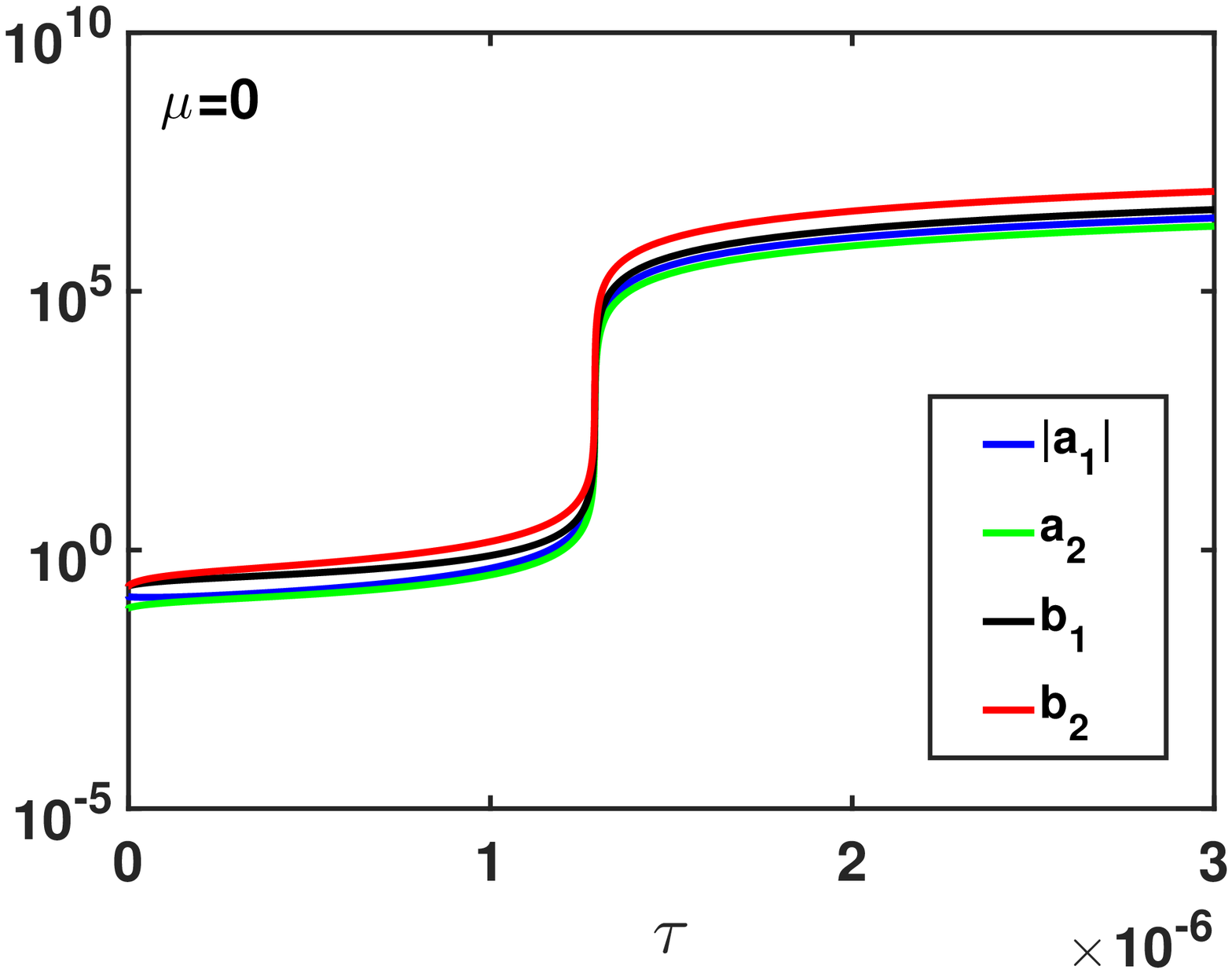}}
  \\
  \subfigure[]
  {\label{2c}
  \includegraphics[scale=.35]{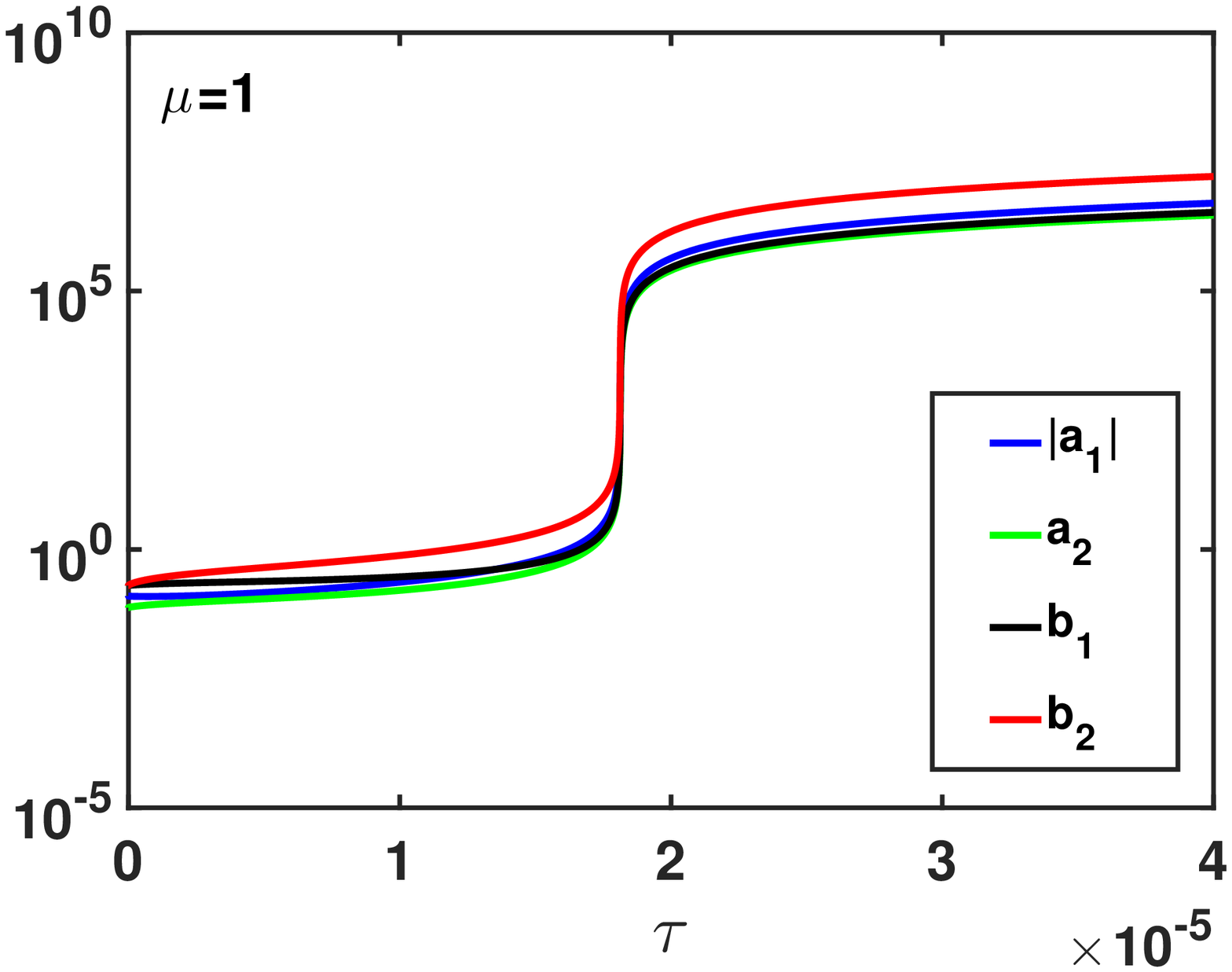}}
  \hskip-.6cm
  \subfigure[]
  {\label{2d}
  \includegraphics[scale=.35]{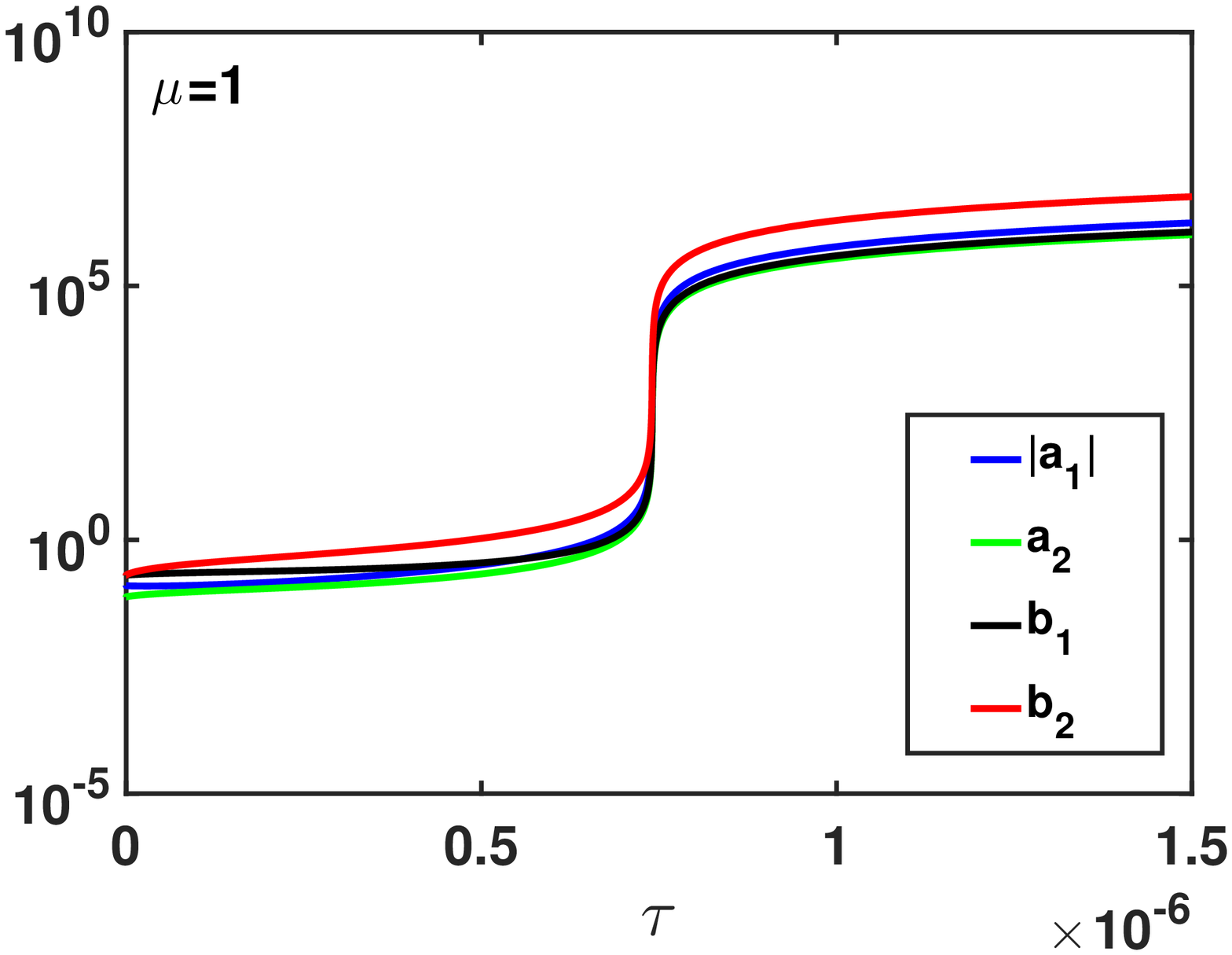}}
  \protect
  \caption{The behavior of the coefficients $a_{1,2}$ and $b_{1,2}$ for different $\mu$
  and $\kappa$ accounting the NS cooling in eq.~(\ref{cooling}).
  (a)~$\mu = 0$ and $\kappa = 10^7$;
  (b)~$\mu = 0$ and $\kappa = 10^8$;
  (c)~$\mu = 1$ and $\kappa = 10^7$;
  (d)~$\mu = 1$ and $\kappa = 10^8$.
  \label{fig:cooling}}
\end{figure}

In this section, we present the results of the numerical solution of eqs.~\eqref{potential} and~\eqref{magfield} accounting for the initial conditions formulated in section~\ref{sec:inicond}. The growth of the magnetic field amplitudes $b_{1,2}(t)$ and the azimuthal potentials $a_{1,2}(t)$ is illustrated below in plots both neglecting the NS cooling in figure~\ref{fig:nocooling} and accounting for such a cooling, accordingly to eq.~(\ref{cooling}), in figure~\ref{fig:cooling}. 

For simplicity, as the first step, we considered above in section~\ref{Sec:saturation} the saturation regime for which the chiral imbalance $|\mu_5^{(\mathrm{sat})}|$ occurs at the level $\sim 3\times 10^{-4}~{\rm eV}\times (B_r/{\rm MeV}^2)$. Thus the corresponding chiral anomaly parameter $\kappa=3.5\times 10^4$ enters differential eqs.~(\ref{potential}) and~(\ref{magfield}), if we consider the common parabolic density  profile suggested in ref.~\cite{Lattimer}. Unfortunately, for such density profile~\cite{Lattimer} and that chiral parameter $\kappa$, the instability could take place during rather late times exceeding the universe age.

\begin{figure}
  \includegraphics[scale=.35]{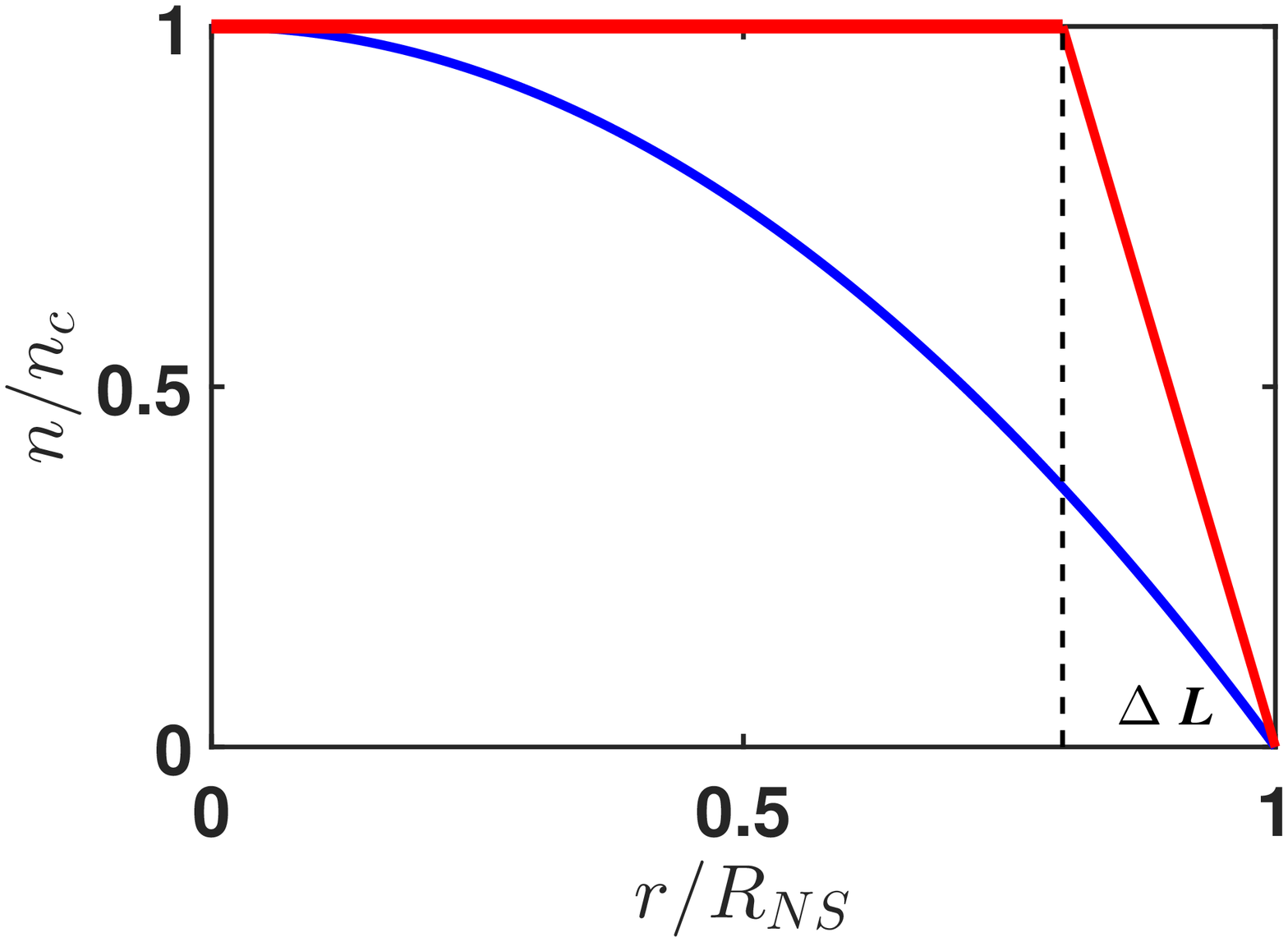}
  \centering
  \protect
  \caption{The schematic illustration of the density distribution in NS. Blue line is the density
  profile proposed in ref.~\cite{Lattimer}. Red line is the simplified density distribution model
  adopted in our work. It corresponds to a uniform core and a thin crust with the depth $\Delta L$, 
  where the density decreases linearly with $r$.
  \label{fig:Delta}}
\end{figure}

In figures~\ref{fig:nocooling} and~\ref{fig:cooling}, we show instead the enhancement, e.g., for $\kappa=10^7$ that can be obtained for a steeper descent of the density profile from the core density characterized by the scale $\Delta L$, see in figure~\ref{fig:Delta}.
For instance, for the model\footnote{See last line in Table~1 in ref.~\cite{Gnedin:2000me}.} with the maximum allowable NS mass $M_\mathrm{NS}=1.73 M_{\odot}$, one can use $\Delta L\sim 350\,\text{m}$  to increase $\kappa$ by three order of magnitude, $\kappa=3.5\times 10^4\to \kappa\times 20\times R_\mathrm{NS}/2\Delta L= 10^7$ that is used in our calculations illustrated in figures~\ref{fig:nocooling} and~\ref{fig:cooling}. Such NS has the central mass density $\rho_{14}^c=32.5$, or number density $n_c=2\times 10^{39}\,\text{cm}^{-3}$, which is twenty times bigger than $10^{38}\,\text{cm}^{-3}$ used above, and the electron chemical potential is  bigger correspondingly, $\mu_e\simeq 250~{\rm MeV}$. Note that the scale $\Delta L=350\,\text{m}$ for density decline is still shorter than the corresponding crust width for such superdense NS, $\Delta R_\mathrm{crust} = 470\,\text{m}$~\cite{Gnedin:2000me}.

One can see in figure~\ref{fig:nocooling}, where one neglects the NS cooling, that after a jump at the critical time moments $t_\mathrm{critical}=t_\mathrm{diff}\tau_\mathrm{critical}\simeq 10^5\,\text{yr}$ for $\kappa=10^7$ and $t_\mathrm{critical}=t_\mathrm{diff}\tau_\mathrm{critical}\simeq 10^4\,\text{yr}$ for $\kappa=10^8$, where $t_\mathrm{diff}=\sigma_0 R_\mathrm{NS}^2=5\times 10^{11}\,\text{yr}$, the magnetic field reaches huge strengths $B\sim 5\times 10^{18}\,\text{G}$,  rising on more 5 orders from the initial $B_i=0.2~\text{MeV}^2=10^{13}\,\text{G}$.

The situation  in figure~\ref{fig:cooling}, where the NS cooling given by eq.~(\ref{cooling}) is taken into account, is similar to the case shown in figure~\ref{fig:nocooling}. However, the critical time that corresponds to the neutrino cooling era $t< 10^6\,\text{yr}$ takes place for $\kappa=10^8$ only; cf. figures~\ref{2b} and~\ref{2d}. For $\kappa=10^7$, justified above for a steep density profile in a superdense NS with the mass $M_\mathrm{NS}=1.73 M_{\odot}$~\cite{Gnedin:2000me}, the critical time $t_\mathrm{critical}\sim 10^7\,\text{yr}$ in figures~\ref{2a} and~\ref{2c} corresponds rather to the photon cooling era.

The dependence on the width parameter $\mu$ for a thin layer just beneath the NS crust is not very essential: for $\mu=1$ the critical time $\tau_\mathrm{critical}$ is a bit less than for the zero width $\mu=0$. One can see it by comparing figures~\ref{1a}, \ref{1b}, \ref{2a}, and~\ref{2b} with figures~\ref{1c}, \ref{1d}, \ref{2c}, and~\ref{2d}.
As we mentioned in section~\ref{Sec:diffur}, in our numerical calculations, we take into account the quenching factor $(1 + B^2/B_0^2)^{-1}$ in eq.~(\ref{saturation}). Nevertheless, the growth of the magnetic field continues even after $\tau_\mathrm{critical}$. This happens due to a strong nonlinearity of AMHD equations where the anomalous current $j\sim \mu_5(B_r)B$ is also quadratic over $B$.

\section{Discussion \label{Sec:Discussion}}

In the present work, we have revealed how the CSE given by the mean spin $\mathbf{S}(\mathbf{x},t)$ in eq.~(\ref{CSE}) supports for a long time the CME that is a driver of the magnetic field instability in NS. 

Recently, a similar term in the equation for the chiral imbalance was accounted for in  ref.~\cite{Schober:2018wlo} in the form, $\sim C_5({\bf B}\cdot\nabla)\mu_e$, where $C_5=1$. Note that, such a contribution was omitted in ref.~\cite{Rogachevskii:2017uyc}, where the ABJ anomaly was taken into account in the form, $\partial_{\mu}J^{\mu}_5=2\alpha_{em}({\bf E}\cdot{\bf B})/\pi$, where $J^{\mu}_5=(n_5, {\bf J}_5)$. The application of the chiral MHD to the hot plasma in the early universe, where $\mu_e$ is negligible, was discussed in ref.~\cite{Schober:2018wlo}. The case of a dense matter in NS, considered in our work, is quite different.

For a non-uniform NS core, where $\mathrm{d}n_e(r)/\mathrm{d}r\neq 0$, the mean spin term differs from zero, $\nabla\cdot\mathbf{S}\neq 0$. Then, accounting for the rigid NS rotation that allowed to avoid the involvement of the Navier-Stokes equation for the fluid velocity $\mathbf{v}$, we have derived in section~\ref{Sec:master} the complete system of the evolution equations in AMHD for the chiral imbalance $\mu_5(\mathbf{x},t)$ and the magnetic field $\mathbf{B}(\mathbf{x},t)$ given by eqs.~(\ref{anomaly}) and~(\ref{Faraday}) correspondingly\footnote{In section~\ref{Sec:CVEvsCME}, we have showed that the CVE given by the term ${\bm\omega}=\nabla\times\mathbf{v}$ in eqs.~(\ref{anomaly}) and~(\ref{Faraday}) is negligible as well.}. 

Instead of the solution of these complicated  self-consistent equations, we have considered in section~\ref{Sec:saturation} the saturation regime, $\partial_t\mu_5(\mathbf{x},t)=0$ substituting $\mu_5^{(\mathrm{sat})}(\mathbf{x})$ from eq.~(\ref{saturation}) into the Faraday eq.~(\ref{Faraday1}) that becomes strongly nonlinear contrary to the standard MHD induction equation. Indeed, in AMHD, the chiral imbalance $\mu_5$ itself depends on magnetic field, $\mu_5^{(\mathrm{sat})}\sim B_r$. Hence the anomalous current in the Faraday equation becomes quadratic over the field owing to the pseudo-scalar spin source $\nabla\cdot\mathbf{S}$ that feeds the chiral anomaly, $\mathbf{j}_\mathrm{anom}\sim B_r\mathbf{B}$. 

Of course, the substitution of the \textit{stationary} chiral anomaly in eq.~(\ref{saturation}) into eq.~(\ref{Faraday1}) is not fully justified. Nevertheless, this simplification allowed us to give some estimates for a strong amplification of magnetic fields in NS. For example, we have obtained the typical time scales for the magnetic field growth in section~\ref{Sec:alpha_squared} using the $\alpha^2$-dynamo for the magnetic field depending on one spatial coordinate, as well as in section~\ref{Sec:Parker} using the mean field dynamo model for 3D axisymmetric magnetic field in NS. The amplification starts from the initial field $B_i\simeq 10^{13}\,\text{G}$ and rises up to the strength $B\sim 10^{18}\,\text{G}$ accounting for the NS cooling through eq.~(\ref{cooling}) at the neutrino cooling era. 

Note that for such superstrong magnetic fields and the chemical potential $\mu_e\simeq 250\,\text{MeV}$ discussed in section~\ref{Sec:Parker1}, a danger equality arises, $2eB\sim \mu_e^2=3.1\times 10^{18}\,\text{G}$,  when our assumption $n_e=\mu_e^3/3\pi^2$ used above, fails in the present approach, which is valid for the case $2eB\ll \mu_e^2$ only. We remind that the opposite inequality $2eB\gg \mu_e^2$ means that all electrons populate the main Landau level $n=0$, where $n_e=n_0=eB\mu_e/2\pi^2$, see in refs.~\cite{Dvornikov:2018tsi,Nunokawa:1997dp}. The latter case inevitably happens for a strong magnetic field which penetrates the NS crust where degenerate electrons are non-relativistic, $p_{F_e}\ll m_e$, $\mu_e=m_e + p_{F_e}^2/2m_e$. However, such a problem is beyond of scope of the present work. 
Hence, at the present moment, we can not  interpret this strong enhancement of the magnetic field up to $B\sim 10^{18}~\text{G}$ postponing for future a solution of this problem. 

Nevertheless, all approximations including the definition of the standard electron density $n_e=\mu_e^3/3\pi^2$, that obeys the necessary condition for strong fields $m_e^2\ll eB\ll \mu_e^2$, remain valid up to the sharp jumps in plots.

Thus, we have revealed how the mean spin source $\nabla\cdot\mathbf{S}$ provided by the polarization of the ultrarelativistic magnetized electron gas in a neutron star produces permanently the chiral imbalance for right and left-handed electrons, $\mu_5=(\mu_\mathrm{R} - \mu_\mathrm{L})/2\neq 0$ that, in its turn, leads to the amplification of a seed magnetic field through the CME.

\acknowledgments


We are thankful to D.~Sokoloff, N.~Kleeorin, and L.~Leinson for helpful discussions.
The work of M.~Dvornikov is supported by the Russian Science Foundation (Grant No.~19-12-00042).

\end{document}